\documentstyle[12pt,a4wide]{article}

\newcommand{\be}[1]{\begin{equation}\label{#1}}
\newcommand{\ee}{\end{equation}}
\newcommand{\ba}[1]{\begin{eqnarray}\label{#1}}
\newcommand{\ea}{\end{eqnarray}}
\begin{document}

\bigskip

\bigskip


\centerline{\Large\bf ON SPONTANEOUS COMPACTIFICATION}
\bigskip
\centerline{\Large\bf IN MULTIDIMENSIONAL COSMOLOGY}

\bigskip

\bigskip
\centerline{ \bf U. BLEYER
\footnote{This work was supported  WIP grant 016659}
}
\centerline{WIP-Gravitationsprojekt, Universit\"at Potsdam }
\centerline{An der Sternwarte 16 }
\centerline{ D-14482 Potsdam, Germany}
\bigskip
\centerline{and}
\bigskip
\centerline{\bf A. ZHUK
\footnote{Permanent address: Department of Physics,
University of Odessa,
2 Petra Velikogo,
Odessa 270100, Ukraine}
\footnote{This work was supported in part by DAAD and by DFG grant 436
UKR - 17/7/93}
}
\centerline{WIP-Gravitationsprojekt, Universit\"at Potsdam }
\centerline{An der Sternwarte 16 }
\centerline{ D-14482 Potsdam, Germany}
\centerline{and}
\centerline{Fachbereich Physik, Freie Universit\"at Berlin}
\centerline{Arnimallee 14}
\centerline{ D-14195 , Germany}

\bigskip

\bigskip

\noindent
{\large\bf Abstract}
\bigskip

\noindent
Mutidimensional cosmological models with $n\left( n\geq 2\right) $
Einstein spaces $M_i\left( i=1,\ldots ,n\right) $ are investigated.
The cosmological constant and homogeneous minimally coupled scalar
field as a matter sources are considered. The scalar field has a
potential of general form depending on the scalar field as well as on
the scale factors of $M_i$ . The general condition of existence of the
solutions with spontaneous compactification is obtained and applied to
physically important cases. Two particular kinds of solutions are
found. One of them has a wormhole-type continuation into the Euclidean
region. Another one represents the generalization of the de Sitter
universe to the multidimensional model under consideration.

\bigskip

\newpage

\section{Introduction}

One of the main problems in multidimensional  cosmology is the
compactification of  extra dimensions. There are two possibilities to
solve it. One of them is the dynamical compactification of extra
dimensions, when the size of the internal dimensions becomes much
smaller than the size of our external dimensions during the evolution
of the universe. Another possibility consists in the proposal that all
extra dimensions are static from the very beginning and they have size
of Planck's length $L_{pl}\sim 10^{-33}$cm to make them unobservable
at present time. This type of solutions is called spontaneous
compactification. There are a lot of papers devoted to
multidimensional cosmological models with spontaneous compactification
[1-13] where this kind of compactification was supposed ad hoc or
where it was obtained solving the equations of motion. In most of
these papers a two-component model was investigated. This means two
factor spaces of constant curvature where one of them is considered as
our external space and the other one is the internal space. Different
kinds of matter were taken as matter sources. But the two-component
picture does not give the possibility to understand the general
situation with spontaneous compactification. All these solutions with
compactification seem to be rather occasional and not connected with
each other. That is why  we consider firstly in this paper {\it
$n$-component models containing $n$ spaces of constant curvature}. In
this case we already obtained few examples of solutions with
spontaneous compactification [14,15]. But they also look rather
occasional  being obtained for particular choices of matter sources
and curvatures of inner spaces. Thus, secondly, we shall investigate
the {\it general model with arbitrary curvatures of inner spaces and a
very general form of the potential}. The main aim of our paper is to
get the {\it conditions for the existence of solutions with
spontaneous compactification for this general model} and to apply
these conditions to different famous cases, for example, to models
with the monopole ansatz by Freund and Rubin [16] or with one-loop
quantum corrections to the potential in compact spaces (Casimir
effect) [5].

The paper is organized as follows. In Section 2. we describe our model
and derive a master equation. This will be discussed for the case of a
free scalar field in Section 3. In Section 4. a nonvanishing potential
will be included into consideration. In Section 5. models with
special potentials depending only on scale factors will be
discussed. In the conclusions  we collect our results. Two particular
kinds of solutions with spontaneous compactification are presented in
Appendices A and B.  References complete the paper.

\section{Model and Master Equation}

In connection with the problem of spontaneous compactification we consider
the cosmological model which is the most natural generalization of the
Friedmann-Robertson-Walker (FRW) metric to multidimensional space with
non-trivial topology: $M_1\times \cdots \times M_n$ , where $M_i$ are
$d _i
$-dimensional spaces of constant curvature.

The metric of this space-time has the form
\be{1}
g =-e^{2\gamma (t)}dt \otimes dt+\sum\limits_{i=1}^na_i^2\left( t\right)
g_{(i)} ,
\end{equation}
where $a_i=e^{\beta ^i}$ denote the scale factors of the factor spaces $M_i$
with the metric $g_{(i)}$ . The Ricci-tensor of $M_i$ is normalized in such
a way that
\be{2}
R_{m_in_i}\left[ g_{(i)}\right] =k_i\left( d_i-1\right) g_{(i)m_in_i}\
,\qquad m_i,n_i=1,\ldots ,d_i
\end{equation}
where $k_i=\pm 1,0$ . Then the scalar curvature of $M_i$ is
\be{3}
R\left[ g_{(i)}\right] \equiv \theta _i=k_id_i\left( d_i-1\right) \ .
\end{equation}
This model can be trivially generalized to Einstein spaces $M_i$ ($i =
1, \dots, n$) for which $R_{m_in_i}[g_{(i)}] =
\lambda_ig_{(i)m_in_i} (\lambda_i$ are arbitrary constants) instead
of (\ref{2}). The results of this paper are valid for Einstein spaces
$M_i$ with the replacement $\theta \rightarrow d_i\lambda_i$ in all
expressions. With this replacement the theorems hold for $M_i$ being
Einstein spaces.

We investigate here the general model with non-zero cosmological constant $
\Lambda $ and homogeneous minimally coupled scalar field $\varphi \left(
t\right) $ with a potential $\tilde V(\beta ,\varphi )$ . This kind of matter
is interesting not only for it's own,
but also because the action for the models with axionic (higher-rank
tensor) fields in homogeneous spaces may be reduced to the action for
the homogeneous minimally coupled free
scalar field [17-19]. For example, it was shown that axionic wormholes
[17] and wormholes with a minimally coupled free scalar field
[14,20,21] are the same (see also the paper by Brown et.al. in [19]).
Higher-rank tensor fields arise also in supergravity
theories. Further, the inclusion of the general potential $\tilde V(\beta
,\varphi )$ depending on the scalar field $\varphi $ as well as on
the scale factors $\beta ^i$ gives us the possibility to investigate theories
with an arbitrary scalar field potential $\tilde V(\beta ,\varphi )\equiv
U(\varphi )$ and, putting $\varphi =const$, we can consider
theories with an arbitrary potential $\tilde V(\beta ,\varphi )\equiv \tilde
V(\beta )$ . Among them are such important models like those
with monopole ansatz [16] as well
as one-loop quantum corrections in compact spaces
(Casimir effect) [5]. At last, the general form of
the potential $\tilde V(\beta ,\varphi )$ gives the possibility to get new
integrable theories. Examples with these kinds of potentials will be
presented here.

Due to the main aim of our paper we look for an
existence condition for solutions with spontaneous compactification
in the framework of the  general model described above, and apply
this condition to the most interesting
known theories. We also find new integrable models which satisfy this
condition. Our paper is a generalization of Maeda's article [9],
where the main attention was paid to the problem of stability of
spontaneously compactified solutions in higher-dimensional theories on
example of two-component models $(n=2)$ without scalar field but with
'monopole' and 'Casimir' form of the potential $\tilde V(\beta )$ . Here we do
not investigate the problem of stability but find the conditions  for
spontaneous compactification in the general model described above.

The action for this model in harmonic-time gauge $\left( \gamma
=\sum\limits_1^nd_i\beta ^i\right) $ has the form [22]
\be{4}
S=\int L\,d\tau ,
\end{equation}
where
\be{5}
L=\frac 12\left( \sum\limits_{i,j=1}^{n}G_{ij}\dot \beta ^i\dot \beta ^j+\dot
\varphi ^2\right) -V .
\end{equation}
The dot denotes differentiation with respect to the harmonic time
$\tau$. The metric of \mbox{minisuperspace} reads
\be{6}
G_{ij}=d_i\delta _{ij}-d_id_j
\end{equation}
and the potential is given by
\be{7}
V=e^{2\sum\limits_{i=1}^{n}d_i\beta ^i}\left[ -\frac
12\sum\limits_{i=1}^{n}\theta
_{i\,}e^{-2\beta ^i}+\tilde V(\beta ,\varphi )+\Lambda \right] \ .
\end{equation}
The equations of motion are
$$
-d_i\ddot \beta ^i+d_i\sum\limits_{k=1}^{n}d_k\ddot \beta
^k+\exp\left(2\sum\limits_{k=1}^{n}d_k\beta ^k\right)\left[
\left( d_i-1\right) \theta_ie^{-2\beta ^i}+
d_i\sum\limits_{k\neq i}\theta _{k\,}e^{-2\beta ^k}
-\right.
$$
\be{8}
\left. -\frac{\partial \tilde V}{\partial \beta ^i}
-2d_i\tilde V-2d_i\Lambda \right] =0\ ,\qquad i=1,\ldots ,n ,
\end{equation}
\be{9}
\ddot \varphi \,+\frac{\partial \tilde V}{\partial \varphi }
\exp\left(2\sum\limits_{i=1}^{n}d_i\beta ^i\right)=0  .
\end{equation}
The constraint reads
\be{10}
\frac 12\left( \sum\limits_{i,j=1}^{n}G_{ij}\dot \beta ^i\dot \beta ^j+
\dot \varphi^2\right) +V=0 .
\end{equation}

Now, one of the spaces $M_i$ , let it be $M_1$ , is considered as our external
space undergoing  dynamical evolution. The other $M_{i\ }($ $i=2,\ldots ,n)$
are internal spaces. The solutions with spontaneous compactification
correspond to the static internal spaces. It means that the scale factors
$a_i=e^{\beta ^i}\ (i=2,\ldots ,n)$ should be constant. Taking
$\beta ^i=const\ (i=2,\ldots ,n)$ in equation (8) we get $n$ dynamical
equations for $\beta ^1$:
$$
\ddot \beta ^1=-\exp\left(2\sum_{k=1}^nd_k\beta ^k\right)\left[
\frac{\theta _1}{d_1}
e^{-2\beta ^1}+\frac 1{d_1-1}\sum_{k=2}^n\theta _k\,e^{-2\beta
^k}- \right.
$$
\be{11}
\left.-\frac 1{d_1(d_1-1)}\frac{\partial \tilde V}{\partial \beta ^1}
-\frac{2\left( \tilde V+\Lambda \right) }{d_1-1}\right] \ ,
\end{equation}
$$
\ddot \beta ^1=-\exp\left(2\sum_{k=1}^nd_k\beta ^k\right)
\left[ \frac{\theta _1}{d_1}
e^{-2\beta ^1}+\frac{\left( d_i-1\right) \theta _i}{d_id_1}e^{-2\beta
^i}+\frac 1{d_1}\sum\limits_{k\neq i}\theta _{k\,}e^{-2\beta ^k}-\right.
$$
\begin{equation}
\left. -\frac 1{d_id_1}\frac{\partial \tilde V}{\partial \beta ^i}
-\frac{2\left( \tilde V+\Lambda \right) }{d_1}\right] \ ,\qquad
i=2,\ldots ,n .
\end{equation}
These equations should be compatible. It is easy to see that this condition
leads to the following theorem:
\\ THEOREM I:\\
{\em The cosmological models with $n\ (n\geq 2)$ spaces $M_i$
$(i=1,\ldots ,n)$of
constant curvatures $\theta _i$, the cosmological constant $\Lambda $ and
homogeneous minimally coupled scalar field with potential $\tilde V(\beta
,\varphi )$ possess solutions with spontaneous compactification of inner
spaces $M_i\ (i=2,\ldots ,n)$ , i.e. $a_i=e^{\beta ^i}=const\ $ $(i=2,\ldots
,n)\ $,only if they satisfy the condition}
\be{13}
\left. \left( \sum\limits_{k=1}^n d_k-1\right) \left( \frac{\theta _i}{d_i}%
e^{-2\beta ^i}+\frac 1{d_i}\frac{\partial \tilde V}{\partial \beta ^i}%
\right) =\sum\limits_{k=1}^{n}\frac{\partial \tilde V}{\partial \beta ^k}
+2\left(
\tilde V+\Lambda \right) \right| _{\beta ^2,\ldots ,\beta ^n=const} ,
\end{equation}
$$
i=2,\ldots ,n .
$$
This is our {\ {\em Master equation}}. From this equation we can derive
two other useful relationships:
\be{14}
\left. \left( \sum\limits_{k=1}^nd_k-1\right) \frac{\theta _i}{d_i}e^{-2\beta
^i}-2\Lambda =-\frac{\left( \sum\limits_{k=1}^nd_k-1\right) }{d_i}\frac{
\partial
\tilde V}{\partial \beta ^i}+\sum\limits_{k=1}^n\frac{\partial \tilde V}{
\partial \beta ^k}+2\tilde V\right| _{\beta ^2,\ldots ,\beta ^n=const}\equiv
A_{(1)i}  ,
\end{equation}
$$
i=2,\ldots ,n .
$$
and
\be{15}
\left. \frac{\theta _i}{d_i}e^{-2\beta ^i}-\frac{\theta _k}{d_k}e^{-2\beta
^k}=-\frac 1{d_i}\frac{\partial \tilde V}{\partial \beta ^i}+\frac 1{d_k}
\frac{\partial \tilde V}{\partial \beta ^k}\right| _{\beta ^2,\ldots ,\beta
^n=const}\equiv A_{(2)ik} ,
\end{equation}
$$
i,k=2,\ldots ,n .
$$
The left-hand sides of (14) and (15) are functions of $\beta ^2,\ldots
,\beta ^n$ and are constants if $\beta ^2,\ldots ,\beta ^n=const$. Thus, in
this case the expressions $A_{(1)i}$ and $A_{(2)ik}$ are functions of $
\beta ^2,\ldots ,\beta ^n$ and are constants as well.
The equations (15) may be rewritten in the following form
\be{16}
\left. \frac{\theta _i}{d_i}e^{-2\beta ^i}+\frac 1{d_i}\frac{\partial \tilde
V}{\partial \beta ^i}=\frac{\theta _k}{d_k}e^{-2\beta ^k}+\frac 1{d_k}\frac{%
\partial \tilde V}{\partial \beta ^k}\right| _{\beta ^2,\ldots ,\beta
^n=const}\equiv F .
\end{equation}
The function $F$ may depend on $\beta ^1$ and $\varphi $ and, thus, be a
function of time.

It is clear that the THEOREM I gives the necessary condition for
spontaneous compactification in our model. But the demand that the
parameters of the universe satisfy condition (13) does, in general, not
lead to the static case $(\dot \beta ^i=0\,,\quad i=2,\ldots ,n)$. To show
this we put (13) without the condition $\beta ^2,\ldots ,\beta
^n=const$ into equation (8) and get $(n-1)$ equations
\be{17}
\frac{\sum\limits_{k=2}^nG_{1k}\ddot \beta ^k}{G_{11}}=\frac{%
\sum\limits_{k=2}^nG_{ik}\ddot \beta ^k}{G_{i1}}\ ,\qquad i=2,\ldots
,n .
\end{equation}
This gives us the equations
\be{18}
\ddot \beta ^i=0\ ,\qquad i=2,\ldots ,n .
\end{equation}
with the solutions $\beta ^i=\nu _it+c_i$ , where $\nu_i, c_i$ are
constants of integration. The static case is contained
in this set and corresponds to the special choice of constants of
integration $\nu_i=0$. For $\dot\beta ^i \neq 0\  (i=2,\ldots , n)$
we get from equation (13) the condition
\be{18a}
\frac{\left( \sum\limits_{k=1}^nd_k-1\right) }{d_i}\left(\theta _ie^{-2\beta^i}
+\frac{\partial\tilde V}{\partial \beta ^i}\right)
-\sum\limits_{k=1}^n\frac{\partial \tilde V}{\partial \beta ^k}
-2\tilde V = 2\Lambda = \mbox{const} .
\end{equation}
and this leads to very strong restrictions to the models. Nevertheless,
examples of such models have been presented in [15,23] and these
models have had solutions with spontaneous compactification as a
particular case.

Let us consider in more detail the condition (13) for some particular forms
of the potential $\tilde V$ .

\section{Free Scalar Field: $\tilde V\equiv 0$}

In this case from the master equation we can get the relations
\be{19}
\frac{\theta _i}{d_i}e^{-2\beta ^i}=\frac{\theta _k}{d_k}e^{-2\beta ^k}=
\frac{2\Lambda }{\sum\limits_{k=1}^nd_k-1}\ ,\qquad i,k=2,\ldots ,n .
\end{equation}

The conditions (\ref{19}) are trivially satisfied  for arbitrary fixed scale
factors $a_{(0)i}\ (i=2,\ldots ,n)$ if all inner factor spaces are
Ricci flat and the cosmological constant is absent: $\theta _2=\ldots
=\theta _n=\Lambda=0$ . If the inner factor spaces are not Ricci flat
the cosmological constant should be present and the fine tuning
(\ref{19}) should take place. In this case all the inner factor spaces
should be non Ricci flat and $\mbox{sign}\theta _i=\mbox{sign}\theta
_k=\mbox{sign}\Lambda \ ,$ $(i=2,\ldots ,n)$.

It is easy to see that
the content of this paragraph will not be changed if no free
scalar field is present at all. This case we shall call the vacuum one.
Thus we have the following
\\ THEOREM II:
{\em Cosmological models with $n(n\geq 2)$ spaces of constant
curvature and a cosmological constant possess only two kinds of solutions
with spontaneous compactification for the case a  homogeneous
minimally coupled free scalar field is present as well as in the
vacuum case . One type of solutions exists for arbitrary fixed scale
factors $a_{(0)i}\ (i=2,\ldots ,n)$ of the inner spaces $M_i$ if all
of them are Ricci flat and the cosmological constant is absent:
$\theta _2=\ldots =\theta _n=\Lambda =0$ . The other type of solutions
corresponds to the fine tuning: $\theta _i/\left( d_ia_{(0)i}^2\right)
=\theta _k/\left( d_ka_{(0)k}^2\right) =2\Lambda /\left(
\sum\limits_1^nd_k-1\right) \ ,\quad i,k=2,\ldots ,n\,.$}

It is clear that the models considered in this chapter satisfy condition
(\ref{18a}) only if $\theta_2= \dots =\theta_n=\Lambda=0$ \cite{15,23}.

It is interesting to note that the fine tuning (\ref{19}) represents
the application of Einstein's idea about a static universe to the
multidimensional case. For models considered in this paragraph (and
for some models with non vanishing potential $\tilde V$, as we shall
see later on) the presence of a non vanishing cosmological constant is
a necessary condition for stabilization of inner compact non Ricci
flat spaces.

\section{Models with Potential $\tilde V(\beta ,\varphi )=U(\varphi )
\widetilde{\tilde V}(\beta )$ }

In what follows we assume that in the presence of a scalar field the
potential $ \tilde V(\beta ,\varphi )$ is factorized, $\tilde V(\beta
,\varphi )=U(\varphi )\widetilde{\tilde V}(\beta )$ . If
$\widetilde{\tilde V}(\beta )\equiv const$, then we obtain the usual
scalar field potential $\tilde V(\beta ,\varphi )\equiv U(\varphi )$ .
Supposing the opposite, $\varphi = $const, we get the case without
scalar field and with potential $\tilde V(\beta ,\varphi )\equiv
\tilde V(\beta )$ .

This paragraph is devoted to the
case $\varphi \neq $ const , i.e. $U(\varphi )$ is a dynamical function. This
fact leads to the demand $A_{(1)i}=A_{(2)in}=0$ in the equations (14) and
(15). It is clear that in the case of a pure scalar field potential $\tilde
V\equiv U(\varphi )$ this demand is not satisfied. Thus, we have
\\ THEOREM III: \\
{\em Cosmological models with $n(n\geq 2)$ spaces of constant
curvature in the presence or absence of the
cosmological constant and with a homogeneous minimally coupled
scalar field never have solutions with spontaneous compactification of
the inner spaces if the scalar field has a non vanishing potential
$\tilde V\equiv U(\varphi )$}.

Let us continue the investigation of the potential $\tilde V=U(\varphi
) \widetilde{\tilde V}(\beta )$ . The demand $A_{(1)i}=A_{(2)ik}=0$ in
equations (14) and (15) leads again to the condition (\ref{19}) for
spontaneous compactification. Thus, either all inner spaces are Ricci
flat and the cosmological constant is absent, $\theta _2=\ldots
=\theta _n=\Lambda =0$ for arbitrary $a_{(0)i}\quad (i=2,\ldots ,n)$ ,
or all inner spaces are not Ricci flat and the fine tuning (\ref{19})
should take place. For both of these types of spontaneous
compactification the potential $\widetilde{\tilde V}$ should satisfy
the equations which follow from (14), (15) and from the demand
$A_{(1)i}=A_{(2)ik}=0$ . Thus, we proved the
\\ THEOREM IV:\\
{\em Cosmological models with $n(n\geq 2)$ spaces of constant
curvature, the cosmological constant,  and a homogeneous minimally
coupled scalar field with potential $\tilde V(\beta ,\varphi
)=U(\varphi )\widetilde{\tilde V} (\beta )$ possess only two kinds of
solutions with spontaneous compactification. One type of solutions
exists for arbitrary fixed scale factors $a_{(0)i}\quad (i=2,\ldots
,n)$ of inner spaces $M_i$ if all of them are Ricci flat and the
cosmological constant is absent, $\theta _2=\ldots =\theta _n=\Lambda
=0$ . The other type of solutions corresponds to the fine tuning
$\theta _i/(d_ia_{(0)i}^2)=\theta _k/(d_ka_{(0)k}^2)=2\Lambda /\left(
\sum\limits_{k=1}^nd_k-1\right) ,\ i,k=2,\ldots ,n$ . In both of these cases
the factor potential $\widetilde{\tilde V}(\beta )$ should satisfy the
equations}
\be{20}
-\frac{\left( \sum\limits_{k=1}^nd_k-1\right) }{d_i}\frac{\partial \widetilde
{\tilde V}}{\partial \beta ^i}+\sum\limits_1^n\frac{\partial \widetilde
{\tilde V}}{\partial \beta ^k}+2\widetilde{\tilde V}=0\ ,\qquad
i=2,\ldots ,n .
\end{equation}
{\em and}
\be{21}
\frac 1{d_i}\frac{\partial \widetilde{\tilde V}}{\partial \beta ^i}=\frac
1{d_k}\frac{\partial \widetilde{\tilde V}}{\partial \beta ^k}\ ,\qquad
i,k=2,\ldots ,n .
\end{equation}

The equation (\ref{20}) shows that the models satisfying
\mbox{THEOREM IV} belong also to the class of models satisfying the
condition (\ref{18a}) if only $\theta _2=\ldots=\theta _n=\Lambda =0$.

Now, we would like to represent a few examples of the factor potential
$\widetilde{\tilde V}$ satisfying equations (\ref{20}) and (\ref{21}).

a) If we suppose
that the potential $\widetilde{\tilde V}$ has the form $\widetilde{\tilde V}
=f_1\left( \sum\limits_{k=2}^nd_k\beta ^k\right) f_2\left( \beta ^1\right)
+f_3\left( \beta ^1\right) + $const then the equations (\ref{20}) and
(\ref{21}) give the
possibility to find the functions $f_1,f_2$ and $f_3$ and we get
finally
\be{22}
\widetilde{\tilde V}=C_1e^{\frac{2+c}{d_1-1}\sum\limits_{k=2}^nd_k\beta
^k+c\beta ^1}+C_2e^{-2\beta ^1}\ ,
\end{equation}
where $C_1,C_2$ and $c$ are arbitrary numbers. For $C_2=c=0$ we have
\be{23}
\widetilde{\tilde V}=C_1e^{\frac 2{d_1-1}\sum\limits_{k=2}^nd_k\beta ^k}
 = \left.  C_1e^{\sum\limits_{k=2}^nd_k\beta ^k}\right|_{d_1=3} .
\end{equation}
If $c=-2d_1, C_2=0$ we obtain
\be{24}
\widetilde{\tilde V}=C_1e^{-2\sum\limits_{k=1}^nd_k\beta ^k} .
\end{equation}

The potential (\ref{24}) has an important property. We can easily see from the
equations (8)-(10) that in this case the gravitational and scalar parts of
the system are not coupled to each other any more. If the model is
integrable without any scalar field, then the inclusion of a scalar
field with the factor potential (\ref{24}) and, for example, the
harmonic factor potential $U(\varphi )=\frac 12m^2\varphi ^2$ does not
change the integrability condition of the system.

b) If we suppose that the potential $\widetilde{\tilde V}$ has the form $
\widetilde{\tilde V}=\frac 12\sum\limits_{k=2}^n\tilde A_ke^{-2\beta ^k}
+f(\beta^1)+$const where $\tilde A_k$ are arbitrary numbers, then we
get with the help of equations (\ref{20}) and (\ref{21}):
\be{25}
\widetilde{\tilde V}=\frac{\tilde A}2\sum\limits_2^n\theta _ke^{-2\beta ^k}+
\widetilde{\tilde A}e^{-2\beta ^1}+B
\end{equation}
where $\tilde A,\ \widetilde{\tilde A}$ are arbitrary numbers and the
constant $B$ should be fine tuned,
\newline
$B=-\frac{\tilde A}2C_0\left(d_1-1\right)$,
where we introduce the definition
\be{26}
\frac{\theta _i}{d_i}e^{-2\beta ^i}=\frac{\theta _k}{d_k}e^{-2\beta ^k}=
\frac{2\Lambda }{\sum\limits_{k=1}^nd_k-1}\equiv C_0 \qquad i,k = 2,
\dots,n .
\end{equation}
$C_0=0$ corresponds to the case $\theta _2=\ldots =\theta _n=\Lambda =0$ .

Let us turn back to the models which possess solutions with spontaneous
compactification in accordance to THEOREM IV (or THEOREM II for
$\tilde V\equiv 0$). For these models all
dynamical equations (8) are reduced to one for the scale factor
$a_1=e^{\beta ^1}$ and this equation reads
\be{29}
\ddot \beta ^1=-e^{2\sum\limits_{k=1}^nd_k\beta ^k}\left[ \frac{\theta _1}{d_1}
e^{-2\beta ^1}-C_0-\frac 1{d_1-1}\left( \frac 1{d_1}\frac{\partial \tilde V}{
\partial \beta ^1}+2\tilde V\right) \right] ,
\end{equation}
where $C_0$ is defined by (\ref{26}). The constraint (10) has the form
$$
d_1\left( d_1-1\right) \dot \beta ^{1^2}=\dot \varphi ^2-\theta _1e^{2\left(
d_1-1\right) \beta ^1}e^{2\sum\limits_{k=2}^nd_k\beta ^k}+
$$
\be{30}
+e^{2d_1\beta
^1}e^{2\sum\limits_{k=2}^nd_k\beta ^k}\left[ 2\tilde
V+C_0(d_1-1)\right] .
\end{equation}
These equations are useful to obtain new integrable cosmological
models. Examples of new solutions of such type are presented in
Appendices A and B. These integrable models are of interest
because they represent the first known to us solutions with an
arbitrary number of non Ricci flat spaces.
Of course, these models are integrable in the case of
spontaneous compactification of inner spaces.  It seems to us hardly
possible to perform the integration
procedure for models with an arbitrary number of non Ricci flat
spaces in the case with all scale factors having dynamical behaviour.

\section{Models with Potential $\tilde V(\beta ,\varphi )\equiv \tilde
V(\beta )\ $}

In this paragraph we investigate models without a dynamical
homogeneous scalar field being present. It means
that we should put $\varphi = $const in our cosmological model.
The potential $\tilde V(\beta,\varphi) \equiv \tilde V(\beta)$ in
(\ref{7}) may now be of different nature. For example, it may arise
due to quantum fluctuations of different fields in compact space [5]
or it has it's origin in a monopole ansatz [16] and so on.

It is clear that all models considered in this paragraph may be
divided into two classes. The models which belong to the first class
possess spontaneous compactification with fine tuning of the form
(\ref{26}) including the case $\theta _2=\ldots =\theta _n=\Lambda
=0$, where the $a_{(0)i}\ \left( i=2,\ldots ,n\right) $ are arbitrary. In
this case the potential $\tilde V(\beta )$ should satisfy the
equations (\ref{20}) and (\ref{21}) (where we should replace the potential
$\widetilde{\tilde V}$ by $\tilde V$ ) and the dynamical behavior of the
system is defined by equations (\ref{29}) and (\ref{30}). Thus we have
\\ THEOREM V: \\
{\em The cosmological models with $n\left( n\geq 2\right) $ spaces of
constant curvature, cosmological constant and the potential $\tilde
V(\beta )$ as a matter source possess solutions with spontaneous
compactification in the case of arbitrary fixed scale factors
$a_{(0)i}\ (\ i=2,\ldots ,n)$ for
\mbox{ $ \theta _2=\ldots =\theta
_n=\Lambda =0$ } as well as in the case of fine tuning, $\theta
_i/(d_ia_{(0)i}^2)=\theta _k/(d_ka_{(0)k}^2)=2\Lambda /\left(
\sum_{k=1}^nd_k-1\right) ,\quad i,k=2,\ldots ,n$ , only if the
potential $\tilde V(\beta )$ satisfies equations (\ref{20}) and
(\ref{21})}.

Let us discuss now two important cases which arise from the ''monopole'' ansatz
by Freund-Rubin [16] and from one-loop quantum corrections in compact
spaces (''Casimir effect'') [5].
In the first case the potential has form [9,11]
\be{27}
{\tilde V}_{mon}=\sum\limits_{k=2}^n\frac{A_k}{e^{2d_k\beta ^k}}
\end{equation}
where $A_k$ are arbitrary numbers. For the second case we have [9,11]
\be{28}
{\tilde V}_{cas}=e^{-\sum\limits_{k=2}^nd_k\beta ^k}
\sum\limits_{k=2}^n\frac{B_k}{e^{\left( d_1+1\right) \beta ^k}}
\end{equation}
where $B_k$ are constants depending on the field content and on the
topology of the model [5,9,11].
The Casimir effect is known for odd-dimensional inner spaces with the
topology of spheres, but formally we consider here arbitrary dimensions
$d_k$ and arbitrary signs of ${\theta_k}  (k = 2, \dots, n)$ [24],
because we think that in all these cases the general form of the
potential (\ref{28}) would be the same.

It is easy to see that for reasonable cosmological models
the potentials (\ref{27}) and
(\ref{28}) do not satisfy the equations (\ref{20}) and (\ref{21}).
For example, in monopole case (\ref{27}) the equation (\ref{21}) gives
$2A_i/\exp(2d_i\beta^i)=2A_k/\exp(2d_k\beta^k)\equiv K =$const. Then,
in order to satisfy equation (\ref{20}) we have to put either $K=0$ or
$d_1+n-2=0$. The former condition leads to the vacuum case ($\tilde
V\equiv 0$), the latter one is not realistic ($d_1 \ge1, n\ge2$). A
similar consideration is possible for the potential (\ref{28}).

Thus, the potentials
$\tilde V_{mon}(\beta )$ (\ref{27}) and $\tilde V_{cas}(\beta )$
(\ref{28}) do not satisfy the condition of THEOREM V. They constitute
the second class of models with spontaneous compactification of the
type different from (\ref{26}).
In this case we can not put
$A_{(1)i}=A_{(2)ik}=0$ in (14) and (15) and the general analysis
is difficult to perform. In what follows we restrict ourselves to the
particular case $F=0$ in (\ref{16}).

First of all we consider the ''monopole'' ansatz [16] which leads in
our model to
the potential (\ref{27}).
The demand $F=0$ leads to fine tuning for the coefficients $A_k$ :
\be{31}
A_k=\left. \frac{\theta _k}{2d_k}e^{2(d_k-1)\beta ^k}\right| _{\beta
^k=const} , \qquad k=2,\ldots ,n .
\end{equation}
{}From this expression we see that $sign\,A_k=sign\,\theta _k$ . If we have
the monopole contribution in each inner space $M_i\ (i=2,\ldots ,n)$ , then $%
A_i\neq 0$ . In this case we should demand $\theta _i\neq 0
(i=2,\ldots ,n)
$, i.e. all inner spaces are non Ricci flat. From the master equation (13) it
follows that the cosmological constant should be fine tuned as well
\be{32}
\left. \Lambda =\frac 12\sum\limits_{k=2}^n\frac{\theta _k}{d_k}\left(
d_k-1\right) e^{-2\beta ^k}\right| _{\beta ^2,\ldots ,\beta ^n=const}